\documentclass{moriond}

\usepackage{xspace}
\usepackage{upgreek}
\usepackage{overpic}

\def\be{\begin{equation}}
\def\bea{\begin{eqnarray}}
\def\eea{\end{eqnarray}}

\newcommand{\Photo}{\includegraphics[height=35mm]{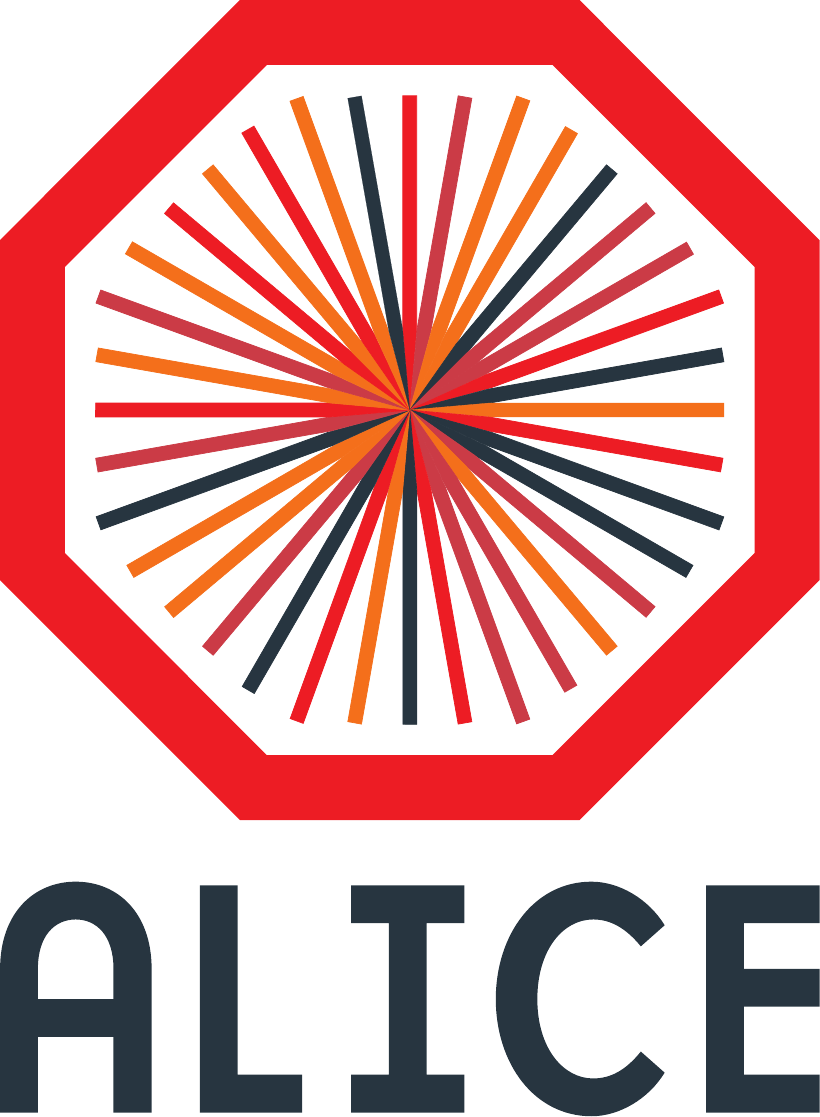}}

\begin{document}
\vspace*{4cm}
\title{Long-range correlations in low-multiplicity pp collisions at $\sqrt{s}=13\,\mathrm{TeV}$}

\author{J.E. Parkkila for the ALICE Collaboration}

\address{CERN, Experimental Physics Department,\\Geneva, Switzerland}

\newcommand{\aac}          {AA\xspace}
\newcommand{\pa}           {pA\xspace}
\newcommand{\pp}           {pp\xspace}
\newcommand{\pplong}       {proton-proton\xspace}
\newcommand{\ppbar}        {\mbox{$\mathrm {p\overline{p}}$}\xspace}
\newcommand{\ee}           {\mbox{$\mathrm {e^{+}e^{-}}$}\xspace}
\newcommand{\ep}           {\mbox{$\mathrm {e^{-}p}$}\xspace}
\newcommand{\PbPb}         {\mbox{Pb--Pb}\xspace}
\newcommand{\AuAu}         {\mbox{Au--Au}\xspace}
\newcommand{\pPb}          {\mbox{p--Pb}\xspace}

\newcommand{\pt}           {\ensuremath{p_{\mathrm{T}}}\xspace}
\newcommand{\kt}           {\ensuremath{k_{\mathrm{T}}}\xspace}
\newcommand{\meanpt}       {$\langle p_{\mathrm{T}}\rangle$\xspace}
\newcommand{\ycms}         {\ensuremath{y_{\rm CMS}}\xspace}
\newcommand{\etarange}[1]  {\mbox{$\left | \eta \right |~<~#1$}\xspace}
\newcommand{\yrange}[1]    {\mbox{$\left | y \right |~<~#1$}\xspace}
\newcommand{\dndy}         {\ensuremath{\mathrm{d}N_\mathrm{ch}/\mathrm{d}y}\xspace}
\newcommand{\dndeta}       {\ensuremath{\mathrm{d}N_\mathrm{ch}/\mathrm{d}\eta}\xspace}
\newcommand{\avdndeta}     {\ensuremath{\langle\dndeta\rangle}\xspace}
\newcommand{\dNdy}         {\ensuremath{\mathrm{d}N_\mathrm{ch}/\mathrm{d}y}\xspace}
\newcommand{\Npart}        {\ensuremath{N_\mathrm{part}}\xspace}
\newcommand{\Ncoll}        {\ensuremath{N_\mathrm{coll}}\xspace}
\newcommand{\dEdx}         {\ensuremath{\mathrm{d}E/\rmfamily{d}x}\xspace}
\newcommand{\RpPb}         {\ensuremath{R_{\rm pPb}}\xspace}
\newcommand{\DeltaM}       {\ensuremath{\Delta M}\xspace}
\let\dNchdeta=\avdndeta
\newcommand{\ntrkl}        {\ensuremath{N_\mathrm{trkl}}\xspace}
\newcommand{\vzeromperc}   {\ensuremath{p_\mathrm{V0M}}\xspace}
\newcommand{\zvtx}         {\ensuremath{z_\mathrm{vtx}}\xspace}
\newcommand{\raa}          {\ensuremath{R_\mathrm{AA}}\xspace}
\let\RAA=\raa
\newcommand{\Ncorrnoave}        {\ensuremath{N_\mathrm{ch}}\xspace}
\newcommand{\Ncorr}        {\ensuremath{\langle N_\mathrm{ch}\rangle}\xspace}

\newcommand{\dEta}           {\ensuremath{\Updelta\eta}}
\newcommand{\dPhi}           {\ensuremath{\Updelta\varphi}}
\newcommand{\Ntrig}          {\ensuremath{N_{\mathrm{trig}}\xspace}}
\newcommand{\Npair}          {\ensuremath{N_{\mathrm{pair}}\xspace}}
\newcommand{\Nassoc}         {\ensuremath{N_{\mathrm{assoc}}\xspace}}
\newcommand{\M}              {\ensuremath{M}\xspace}
\newcommand{\pttrig}         {\ensuremath{p_{\rm T,trig}}\xspace}
\newcommand{\ptassoc}        {\ensuremath{p_{\rm T,assoc}}\xspace}
\newcommand{\vtwo}           {\ensuremath{v_{\rm 2}}\xspace}


\newcommand{\GeV}          {\ensuremath{\mathrm{GeV}}\xspace}
\newcommand{\GeVc}         {\ensuremath{\mathrm{GeV}/c}\xspace}
\newcommand{\TeV}          {\ensuremath{\mathrm{TeV}}\xspace}
\newcommand{\sqrts}        {\ensuremath{\sqrt{s}}\xspace}
\newcommand{\thirteen}     {\ensuremath{\sqrt{s}=13\,\mathrm{TeV}}\xspace}
\newcommand{\nineone}      {\ensuremath{\sqrt{s}=91\,\mathrm{GeV}}\xspace}

\maketitle\abstracts{
In these proceedings, the measurements of the long-range near-side yields of charged hadrons in low-multiplicity proton--proton collisions at \thirteen are presented. The investigation studies pairs of charged particles within $1.4 < |\Delta\eta| < 1.8$ and $1 < \pt < 2\,\GeVc$, exploring the charged-particle multiplicity dependence of these observables at mid-rapidity. The analysis extends the previous studies of two-particle correlations in hadronic collisions by exploring the lowest multiplicity region ever investigated, where the formation of a strongly-interacting medium is less likely. By comparing the results with those obtained in \ee collisions at \nineone, the role of the \ee collision mechanisms in the emergence of collective-like phenomena in small systems can be constrained. The study provides new insights into the origin of long-range correlations in proton--proton collisions. 
}

\section{Introduction}

Long-range correlations are observed in ultra-relativistic heavy-ion collisions at RHIC and LHC, and are generally regarded as evidence for a strongly-coupled medium, quark--gluon plasma (QGP). Similar observations in high-multiplicity proton--proton (pp), proton--nucleus, and light nucleus--nucleus collisions challenge the interpretation of collective phenomena in hadronic collisions. Despite extensive efforts, a clear description of the data has not been achieved, and flow-like signatures may originate from both early and late stages of collisions. Studying such signatures across collision systems and sizes is critical to understanding collective phenomena.

Recently, long-range correlations were studied in \ee collisions at \nineone using archived data from the ALEPH experiment\cite{Badea:2019vey}. No significant long-range correlation was observed. Since the measurements of correlations in electron-positron collisions are not influenced by the presence of beam remnants or initial-state radiations, and are insensitive to parton distribution function modeling\cite{PhysRevC.97.024909,Castorina:2020iia}, they provide a useful comparison to pp collisions, and can help understand the mechanisms of emergence of collective-like signals in small hadronic systems. The results obtained in \ee collisions were also compared to the associated yield measurement in pp collisions with CMS\cite{CMS:2015fgy}, which was rescaled to account for the pseudorapidity acceptance ratio between ALEPH and CMS. Due to the large uncertainty of the existing pp measurement, a statistically significant comparison between the associated yields measured in pp and \ee collisions was not feasible.

In these proceedings, the near-side long-range yield measurements are presented in pp collisions at \thirteen with good accuracy down to very low multiplicities, allowing for a quantitative comparison with \ee collisions at \nineone.

\section{Analysis}
\subsection{Observables}
The two-particle correlation function, which is defined as a function of relative azimuthal angle $\Updelta\varphi$ and pseudorapidity $\Updelta\eta$, can be written as:
\begin{equation}
\label{eq:twocorr}
\frac{1}{N_{\mathrm{trig}}}\frac{\mathrm{d}^{2}N_{\mathrm{pair}}}{\mathrm{d}\Delta\eta\mathrm{d}\Updelta\varphi} = B(0,0)\frac{S(\Updelta\eta,\Updelta\varphi)}{B(\Updelta\eta,\Updelta\varphi)}\Big|_{\pttrig,\,\ptassoc}.
\end{equation}
In Eq.~\ref{eq:twocorr}, $\pttrig$ and $\ptassoc$ represent the transverse momentum of the trigger and associated particles, respectively. $N_\mathrm{trig}$ is the total number of trigger particles, and $N_\mathrm{pair}$ is the number of trigger- and associated particle pairs. The average number of particle pairs originating from a single collision event, $S(\Updelta$$\eta$,$\Updelta$$\varphi)$, is divided by the number of pairs where the associated particle is obtained from a large pool of other but similar mixed events $B(\Updelta\eta$,$\Updelta\varphi)$. This results in the final two-particle correlation function, which is corrected for pair acceptance and reconstruction effects. The mixed-event distribution is normalized by $B(0,0)$ under the assumption of zero acceptance effects, as particles traveling in the same direction see identical efficiencies.

The two-particle correlation function is integrated over two $\Updelta\eta$ intervals, $1.4<|\Delta\eta|<1.8$, to exclude the jet fragmentation peak and obtain a per-trigger yield distribution as a function of $\Updelta\varphi$:
\begin{equation}
\label{eq:pertrigg}
Y(\Updelta\varphi) = \frac{1}{N_{\mathrm{trig}}}\frac{\mathrm{d}N_{\mathrm{pair}}}{\mathrm{d}\Updelta\varphi} = \int_{1.4<|\Updelta\eta|<1.8}\left(\frac{1}{{N}_{\mathrm{trig}}}\frac{\mathrm{d}^{2}N_{\mathrm{pair}}}{\mathrm{d}\Updelta\eta\mathrm{d}\Updelta\varphi}\right) \frac{1}{\delta_{\Updelta\eta}} \mathrm{d}\Updelta\eta-C_\mathrm{ZYAM}.
\end{equation}
This long-range definition sets the normalization constant of the integral to $\delta_{\Updelta\eta}=0.8$, and a Zero-Yield-At-Minimum (ZYAM) procedure is used to subtract the baseline of the correlation function. The $C_\mathrm{ZYAM}$ constant in Eq.~\ref{eq:pertrigg} is determined by fitting a Fourier-series $F(\Updelta\varphi)$ up to the third harmonic to the per-trigger yield distribution. The constant is obtained from the minimum of the Fourier-fit $F(\Updelta\varphi_\mathrm{min})$, minimizing the effect of statistical fluctuations in the correlation. 

The associated ridge yield is then extracted using
\begin{equation}
Y^\mathrm{ridge}=\int_{|\Updelta\varphi|<|\Updelta\varphi_\mathrm{min}|}\frac{1}{N_{\mathrm{trig}}}\frac{\mathrm{d}N_{\mathrm{pair}}}{\mathrm{d}\Updelta\varphi}\mathrm{d}\Updelta\varphi.
\end{equation}
The minimum of the Fourier-fit also provides a reasonable estimate for the integration range $|\Updelta\varphi|<|\Updelta\varphi_\mathrm{min}|$, which can slightly vary from one multiplicity class to another.

\subsection{Estimation of statistical and systematic uncertainties}
Systematic uncertainties for associated ridge yields are assessed through variations in event and track selection, as well as changing the integration range for per-trigger yield extraction in $\Delta\eta$. Generally, the systematic contribution was found small relative to the statistical fluctuations. Bootstrap resampling is used to estimate confidence levels in each multiplicity interval, allowing for consistent uncertainty evaluation across all intervals, as adopted in Ref.\cite{Badea:2019vey}.
For each bin $i$, the central value $y_i$ of the discrete $Y(\Updelta\varphi)$ at $\Updelta\varphi_i$ is modulated by random Gaussian smearing $\delta y_i\sim [1+\mathcal{N}(0,\sigma_{i,\mathrm{stat}}/y_i)+\sum_s(\mathcal{N}(0,R_s-1)+y_{\mathrm{var},i}^s/y_i-R_s)]F(\Updelta\varphi_i)$, where $\mathcal{N}(\mu,\sigma^2)$ is a random sample from a Gaussian distribution, $\sigma_{i,\mathrm{stat}}$ the statistical uncertainty of sample $i$, and $R_s$ is a 0th-order polynomial fit to $y_{\mathrm{var}}^s/y$ with $s$ denoting the systematic variation.
The final uncertainty on $Y^\mathrm{ridge}$ is calculated from the standard deviation of the yield distribution resulting from repeated modulation and extraction.

\begin{figure}[h!]
	\centering
	\hspace{-2em}\begin{overpic}[width=0.9\textwidth]{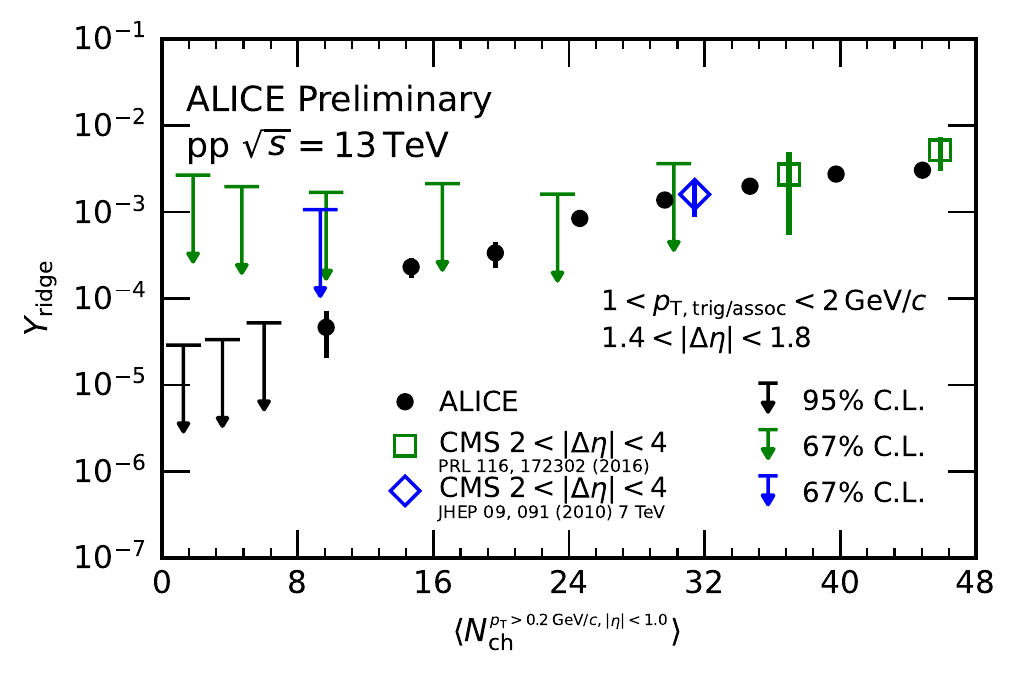}
		\put(0,0){\includegraphics[width=0.36\textwidth]{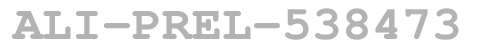}}
	\end{overpic}
	\caption{The extracted associated ridge yield $Y^\mathrm{ridge}$ as a function of the charged-particle multiplicity. The black data points correspond to the measurements by the ALICE, while the green squares represent the CMS measurement. 
	Whenever the lower uncertainty reaches zero, the upper uncertainty limit is represented by a bar and arrow-down for both results. The mean multiplicity of the CMS data points has been adjusted to match the multiplicity definition used in this study by counting the particles within the acceptance region of both experiments using a PYTHIA analysis.}
	\label{fig:alicecms}
%
	\centering
	\hspace{-2em}\begin{overpic}[width=0.9\textwidth]{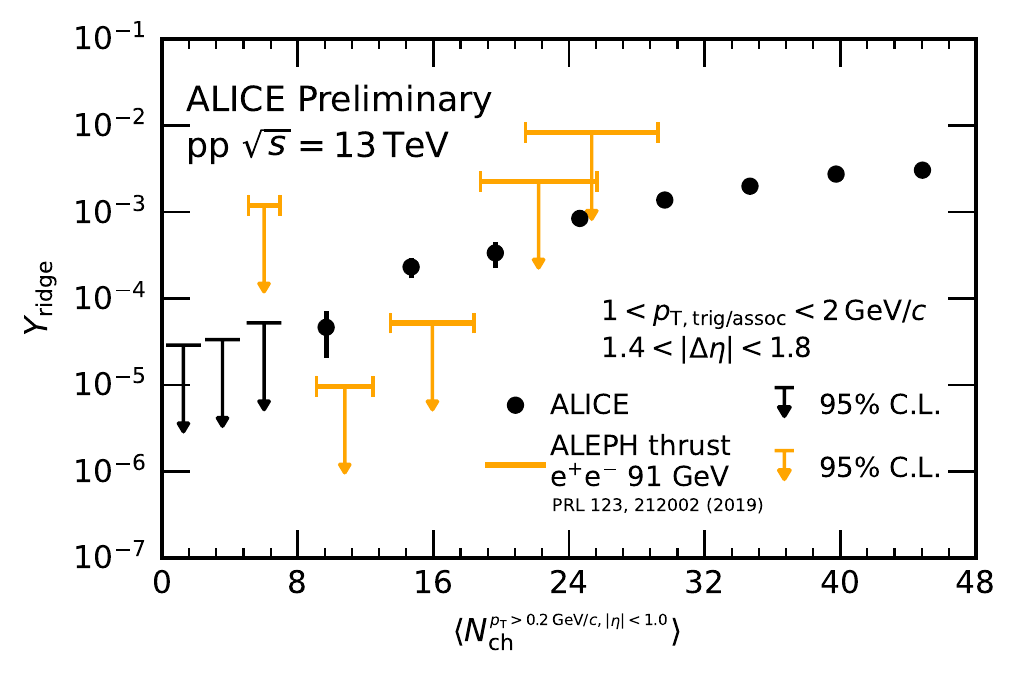}
		\put(0,0){\includegraphics[width=0.36\textwidth]{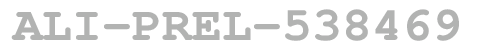}}
	\end{overpic}
	\caption{Ridge yield $Y^\mathrm{ridge}$ as a function of multiplicity. The orange limits indicate the ALEPH \ee thrust coordinate system data points. 
	The horizontal limits in the ALEPH points represent the uncertainty related to the multiplicity conversion from the ALEPH multiplicity definition, given the different collision systems and energy.}
	\label{fig:alicealeph}
\end{figure}

\section{Results}
In Fig.~\ref{fig:alicecms}, the extracted associated ridge yield $Y^\mathrm{ridge}$ in the range $1.0<\pttrig<2.0\,\mathrm{GeV}/c$ is presented as a function of the average charged-particle multiplicity. The measured $Y^\mathrm{ridge}$ shows a strong dependence on the multiplicity with an increasing trend towards higher multiplicity collisions. The measurement extends the low-multiplicity reach of previous experiments\cite{CMS:2015fgy} with good accuracy for events with $\Ncorr$ down to 9.7. 
The confidence limits (C.L.) are presented as black arrows, and the results are compared with the CMS measurement at the same centre-of-mass energy, which presents measured near-side yields for $\Ncorr\geq40$ and C.L. for smaller multiplicities. The comparison shows good agreement at high multiplicities where an accurate estimation of the associated yields is available for both measurements. In Fig.~\ref{fig:alicealeph}, the result is compared to a recent measurement performed in \ee at \nineone using ALEPH archived data\cite{Badea:2019vey}, where no evidence of long-range near-side correlations was observed. The vertical arrows present the estimated C.L., while the horizontal bars indicate the uncertainties associated with the scaling of the x-axis of the ALEPH data. The uncertainty was estimated by simulating PYTHIA events in both pp at \thirteen and \ee at \nineone, and counting the resulting particles in acceptance ranges of both experiments. The ALICE measurements exhibit a larger yield compared to the \ee within $\Ncorr\sim 10$ to $20$ by $3.2\sigma$. Evidently, the processes involved in the \ee annihilations at \nineone are not important in the emergence of collective effects in pp collisions, as the yield measured from an \ee system in this multiplicity range does not reach significant non-zero values.

\section{Conclusions}
These results represent the first accurate measurement of the long-range near-side ridge yield in low-multiplicity pp collisions at \thirteen. The high precision of the measurement allows for quantitative comparisons between the near-side yield in hadronic collisions and \ee annihilations, which is a simpler and better understood system. The results indicate that the yield in hadronic collisions with similar multiplicity is significantly larger than in \ee annihilations, indicating that the processes involved in \ee annihilations may not play a significant role in the production of collectivity in pp collisions. This finding provides valuable insight into the mechanisms of particle production in pp collisions.

\section*{References}

\bibliographystyle{utphys}   
\bibliography{bibliography}

\end{document}